\documentclass[aps,pra,twocolumn,superscriptaddress]{revtex4}

\usepackage{graphicx}
\usepackage{dcolumn}
\usepackage{bm}

\usepackage{amsmath}
\usepackage[latin1]{inputenc}
\usepackage{bm}
\usepackage{graphicx}
\usepackage{SIunits}
\usepackage{subfigure}
\usepackage[inline]{asymptote}
\usepackage{lmodern}
\usepackage{tikz}
\usetikzlibrary{calc}
\usepackage{mwe}

\newcommand{\be}{\begin{equation}}
\newcommand{\ee}{\end{equation}}
\newcommand{\ba}{\begin{eqnarray}}
\newcommand{\ea}{\end{eqnarray}}
\newcommand{\ban}{\begin{eqnarray*}}
\newcommand{\ean}{\end{eqnarray*}}

\newcommand{\braket}[2]{\mbox{$ \langle #1 | #2 \rangle $}}

\newcommand{\ket}[1]{\mbox{$ | #1 \rangle $}}
\newcommand{\bra}[1]{\mbox{$ \langle #1 | $}}

\newcommand*\diff{\mathop{}\!\mathrm{d}}

\begin{document}

\title{Detecting Optical Channel Non-Reciprocity with Non-Local Quantum Geometric Phase}

\author{James E. Troupe}
\affiliation{Applied Research Laboratories, The University of Texas at Austin}
\author{Antia Lamas-Linares}
\affiliation{Texas Advanced Computing Center, The University of Texas at Austin}
\date{\today}
\begin{abstract}
Non-reciprocal devices are of increasing interest in quantum information technologies. This paper examines whether the presence of a non-reciprocal device in an optical channel is detectable by the communicating parties. We find that a non-reciprocal device such as a Faraday Rotator results in a measurable geometric phase for the light propagating through the channel and that, when using entangled photon pairs, the resulting phase is non-local and robust against malicious manipulation.
\end{abstract}
\maketitle
\section{Introduction}

Non reciprocal devices and processes are attracting increasing attention as powerful additional resources in quantum information. This expansion of the quantum optics toolkit used to describe quantum information processes and devices, is referred to as Chiral Quantum Optics, where couplings between quantum systems can depend exclusively on the direction of propagation, leading to additional possibilities for quantum control and applications in quantum technologies~\cite{lodahl:17, muller:18}.

It is in this context that we consider the problem of detecting the non-reciprocity of an optical channel used for the transmission of quantum information encoded in polarization. This study was prompted by a particular application, namely the secure synchronization of remote clocks using entangled photons~\cite{lamas-linares:18}. In that case, a metrological task is believed to be secure under the assumption of a reciprocal channel, which naturally leads to the question of whether or not the non-reciprocity is detectable and whether the detection is robust against malicious intervention. Optical elements that utilize polarization rotation to break reciprocity necessarily induce a geometrical phase on the polarization state of a photon~\cite{kwiat:91}. When this phase is induced on polarization entangled pairs of photons, the phase is non-local~\cite{strekalov:97} and the non-reciprocity detection process is resistant to tampering by an adversary.

\section{Breaking Reciprocity with Circulators}
\begin{figure}
\includegraphics[width=1.0\linewidth]{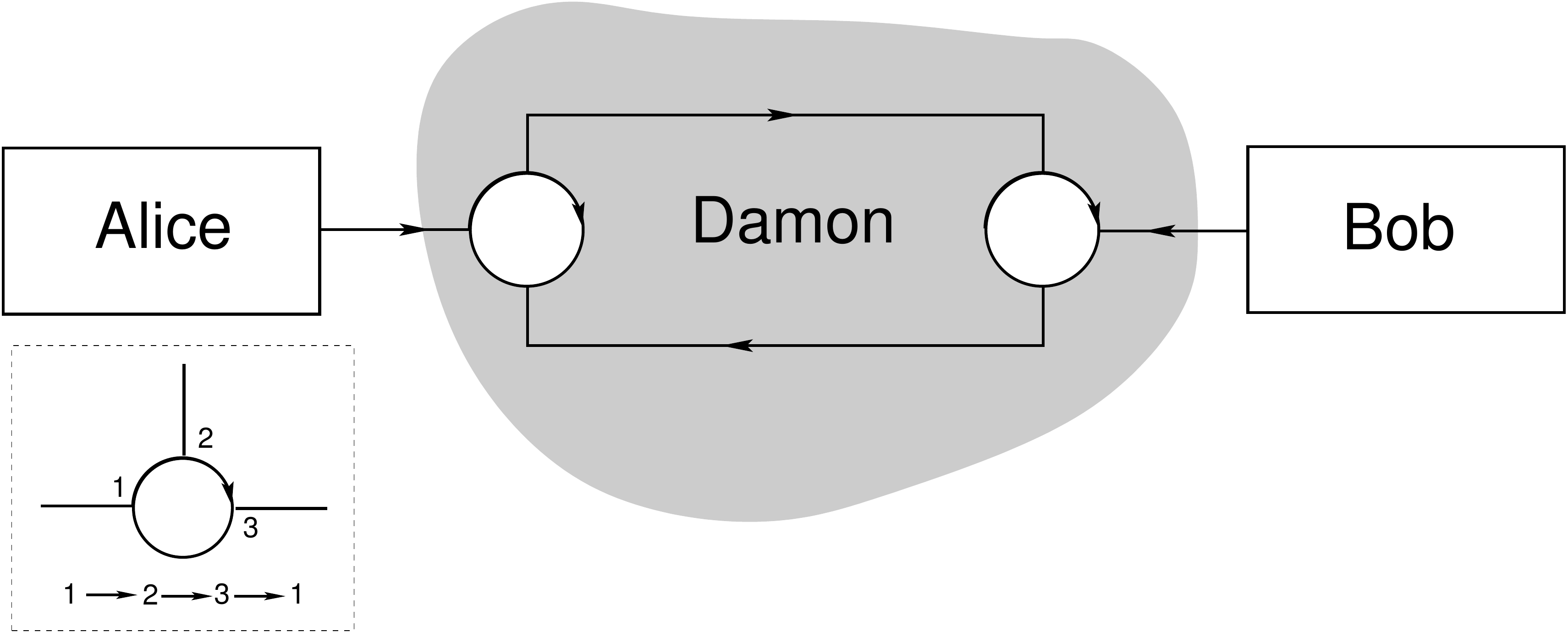}
\caption{Circulator as a non-reciprocal device. A single mode optical channel connects Alice and Bob. The intermediate path is controlled by a malicious adversary Damon, who can use a pair of circulators to break the reciprocity of the channel. In this example we see that photons originating at Bob's site will always take the top path, while photons originating at Alice's side will take the lower path. This allows Damon to introduce an arbitrary asymmetry in the light propagation times between Alice and Bob. The inset shows schematically the action of a circulator with three ports. Light entering port 1 exits through port 2; light entering throuh port 2 exits through port 3 and light entering through port 3 exits through port 1.} \label{fig:circulator}
\end{figure}

An optical channel is reciprocal if the state of the incoming light and the state of the outgoing light are time reversals of each other.  A non-reciprocal optical element, such as a Faraday Rotator (FR), breaks this symmetry.  All passive, linear, and non-magnetic optical elements are reciprocal.  In the case of a channel containing a magnetic optical element such as a Faraday Rotator, if we time-reverse the channel, the direction that the polarization is rotated by the FR does not change. An optical circulator uses a FR to break reciprocity \cite{martinelli:17}, allowing the circulator to passively route light to different ports based on its direction of travel through the device (Fig.~\ref{fig:circulator}) while leaving other degrees of freedom of the light unmodified in any fundamental sense. 

In the context of the clock synchronization procedure described in \cite{lamas-linares:18}, a natural way to break the protocol is to have the propagation time be different for photons moving from Alice to Bob than from Bob to Alice. This is easily achieved by a malicious party (Damon) with the use of polarization independent circulators as shown in Fig.~\ref{fig:circulator}. An important detail is that each circulator rotates the state through a closed loop in the Poincare sphere since the initial and final polarization state must be the same. However the closed path followed by the polarization state as it evolves through the circulator will depend on the starting point. In practice, this means that the polarization state of the photon undergoes a full rotation in a plane defined by the physical configuration of the Faraday Rotators. The combination of two circulators allows Damon to passively introduce a direction dependent delay while preserving the polarization state of all of the individual photons propagating through the channel. If the non-reciprocity of the channel is non-detectable or, more generally, can be compensated for an arbitrary unknown input state, the protocol is broken. If, however, there is a procedure to detect this characteristic of the channel and the detection process cannot be obscured by a malicious adversary, then an assumption of reciprocity of the channel is testable in a security-sensitive context.

In the following section we show that Alice and Bob will always be able to detect any such attack that utilizes non-reciprocal rotation of the photons' polarization. This is due to an unavoidable non-local geometric phase that such a strategy must impart between the entangled photon pairs shared between Alice and Bob.

\section{Geometric phase of a qubit}
As a start, we review how a geometric phase is acquired by the state of a qubit as it is rotated through a closed path on the Bloch sphere \cite{berry:84,aharonov:87}. Let the initial state of the qubit be 
\begin{equation} \label{eq:qubit}
\ket{\psi(t=0)} = e^{-i\phi}\cos(\theta/2)\ket{R}+ \sin(\theta/2)\ket{L}.
\end{equation}
We will define the overall phase of the qubit after time $t$ to be
\begin{equation*}
\ket{\tilde{\psi}(t)}=e^{-if(t)}\ket{\psi(t)}.
\end{equation*}
And we will define the overall phase difference between the qubit state at $t=0$ and $t=\tau$ to be $\Delta f = f(\tau) - f(0)$. 

To derive the evolution of the function $f(t)$ as the qubit rotates about the z-axis, we write  Schr{\"o}dinger's equation for the state: 
\begin{equation}
i\hbar\frac{\diff}{\diff t}\ket{\tilde{\psi}(t)} = i\hbar \left( -i\frac{\diff f}{\diff t} e^{-if(t)}\ket{\psi(t)} + e^{-if(t)} \frac{\diff}{\diff t}\ket{\psi(t)} \right). \nonumber 
\end{equation}
From this, we can see that 
\begin{equation}
\bra{\tilde{\psi}(t)}i\frac{\diff}{\diff t}\ket{\tilde{\psi}(t)} = \frac{\diff f}{\diff t} +  i\langle\tilde{\psi}|e^{-if(t)}\frac{\diff}{\diff t}\ket{\psi(t)}. \nonumber
\end{equation}
And so,
\begin{align}
\frac{\diff f}{\diff t} &= \bra{\tilde{\psi}(t)}i\frac{\diff}{\diff t}\ket{\tilde{\psi}(t)} - \bra{\psi}i\frac{\diff}{\diff t}\ket{\psi(t)} \nonumber \\
&= \bra{\tilde{\psi}(t)}i\frac{\diff }{\diff t}\ket{\tilde{\psi}(t)} - \frac{1}{\hbar}\bra{\psi}\hat{H}\ket{\psi(t)}.  \nonumber
\end{align}
Integrating this over the path taken by the qubit from $t=0$ to $t=\tau$, we have a total phase change $\Delta f$ given by 
\begin{equation}
\Delta f = \int\displaylimits_0^\tau \frac{\diff f}{\diff t}\diff t= \beta - \frac{1}{\hbar}\int\displaylimits_0^\tau \bra{\psi}\hat{H}\ket{\psi(t)} \diff t.
\end{equation}
The first term is the geometric phase and the second term is the phase change due to the qubit's dynamics. 
In our analysis, we will be assuming that the dynamical phase shift is either zero or is known and has been compensated.  The geometric phase is given by 
\begin{equation} \label{eq:GeoPhase}
\beta = \int\displaylimits_0^\tau \bra{\tilde{\psi}(t)}i\frac{\diff}{\diff t}\ket{\tilde{\psi}(t)} \diff t. 
\end{equation}
This term is nonzero because the Bloch sphere has nonzero curvature. The value of the overall phase at any chosen point on the Bloch sphere can be arbitrarily specified; however, the way a chosen phase of must change as the qubit moves on the Bloch sphere is given by the curvature of the state space. 

More formally, the local derivative along the qubit's path on the surface of the Bloch sphere must be invariant under $U(1)$ gauge transformations \cite{simon:83}. The requirement of $U(1)$ invariance is a result of the qubit's state being invariant under changes in the overall phase at each point on the Bloch sphere. To make the local derivative invariant at each point, the usual derivatives must be replaced with covariant derivatives defined by a local gauge field at each point. The covariant derivative essentially defines the notion of parallel transport of a locally defined phase along the qubit's path. The gauge field is defined by $\vec{A}(R(t)):=i\bra{\tilde{\psi}(\vec{R}(t))}\vec{\nabla}_{R}\ket{\tilde{\psi}(\vec{R}(t))}$, where $\vec{R}(t)$ defines the path taken on the Bloch sphere. Equation \ref{eq:GeoPhase} says that the geometric phase can be calculated by integrating the local gauge field along the qubit's path, 
\begin{align}
\beta &= \int\displaylimits_0^\tau i\langle\bra{\tilde{\psi}(\vec{R}(t))}\frac{\diff}{\diff t}\ket{\tilde{\psi}(\vec{R}(t))} \diff t \nonumber \\
&= \oint\displaylimits_{path} \diff \vec{R} \cdot \vec{A}(\vec{R}). \nonumber
\end{align}

If the path is a closed one, then we can use  Stokes' Theorem to convert the path integral to an integral of the curl of the gauge field (the curvature of the Bloch sphere) over the surface enclosed by the path, 
\begin{align}
\beta &= \oint\displaylimits_{path} \diff\vec{R} \cdot \vec{A}(\vec{R}) \nonumber \\
 &= \iint\displaylimits_{surface} \diff\vec{S} \cdot \left( \vec{\nabla} \times \vec{A} \right). \nonumber
\end{align}
Expanding the expression for $\vec{A}$, we see that 
\begin{equation}
\begin{split}
\vec{A} &= i\bra{\tilde{\psi}(\vec{R}(t))}\vec{\nabla}_{R}\ket{\tilde{\psi}(\vec{R}(t))} \\
&= i\bra{\tilde{\psi}(\vec{R}(t))}\vec{\nabla}_{R}\left( e^{-if(\vec{R}(t))} \right)\ket{\psi(\vec{R}(t))} \\
& \quad + i\bra{\psi(\vec{R}(t))}\vec{\nabla}_{R}\ket{\psi(\vec{R}(t))} \\
&=  i\bra{\psi(\vec{R}(t))}\vec{\nabla}_{R}\ket{\psi(\vec{R}(t))} + \vec{\nabla}_{R}f.
\end{split}
\end{equation}
More explicitly, using Equation \ref{eq:qubit}, this becomes 
\begin{align}
\vec{A} &= i\bra{\psi}\frac{\partial}{\partial\theta}\ket{\psi} \hat{\theta} + i\bra{\psi}\frac{1}{\sin\theta}\frac{\partial}{\partial\phi}\ket{\psi} \hat{\phi} + \vec{\nabla}_{R}f \nonumber \\
&= \frac{1}{\sin\theta}\cos^2\left(\theta/2\right) \hat{\phi} + \vec{\nabla}_{R}f. \nonumber
\end{align}
Calculating the curl of the gauge field, we note that the curl of the second term above is zero -- an expression of the gauge invariance of the curvature. Defining the components of $\vec{A}$ to be $A_i$, the curl of $\vec{A}$ on the surface of the Bloch sphere is given by 
\begin{align}
(\vec{\nabla} \times \vec{A})(r=1,\theta,\phi) &= \bigg \{ \frac{1}{r \sin\theta}\left( \frac{\partial}{\partial\theta}\left(A_\phi\sin\theta \right) - \frac{\partial A_\theta }{\partial\phi}\right)\hat{r} \nonumber \\
& \quad + \frac{1}{r} \left( \frac{1}{\sin\theta}\frac{\partial A_r}{\partial\phi} - \frac{\partial A_\phi}{\partial r} \right)\hat{\theta} \nonumber \\
& \quad + \frac{1}{r} \left( \frac{\partial}{\partial r}\left( r A_\theta \right) - \frac{\partial A_r}{\partial\theta}\right)\hat{\phi} \bigg \}_{r=1}\nonumber \\
&= \frac{1}{\sin\theta} \left( \frac{\partial}{\partial\theta} \left( \cos^2(\theta/2) \right) - 0 \right)\hat{r} \nonumber \\
(\vec{\nabla} \times \vec{A})(r=1,\theta,\phi) &= -\frac{1}{2} \hat{r}.  \nonumber
\end{align}
Therefore, we have 
\begin{align}
\beta &= \iint\displaylimits_{surface} \diff \vec{S} \cdot \left( \vec{\nabla} \times \vec{A} \right) \nonumber\\
&= -\frac{1}{2}\iint\displaylimits_{surface} \diff S = -\frac{1}{2} \Omega, \nonumber
\end{align}
where $\Omega$ is the solid angle subtended by the closed path.

In the case of the polarization state of light, this geometric phase was first noted by Pancharatnam \cite{pancharatnam:56} and further elaborated upon by Berry \cite{berry:87}. In the fully quantum context, geometric phase in two-photon interference of entangled photon pairs produced by parametric downconversion has been studied in several contexts \cite{kwiat:91, brendel:95,strekalov:97,hessmo:00, galvez:07,jha:08}. 

\section{Geometric Phase and the Circulator Attack} 
\begin{figure}
\includegraphics[width=1.0\linewidth]{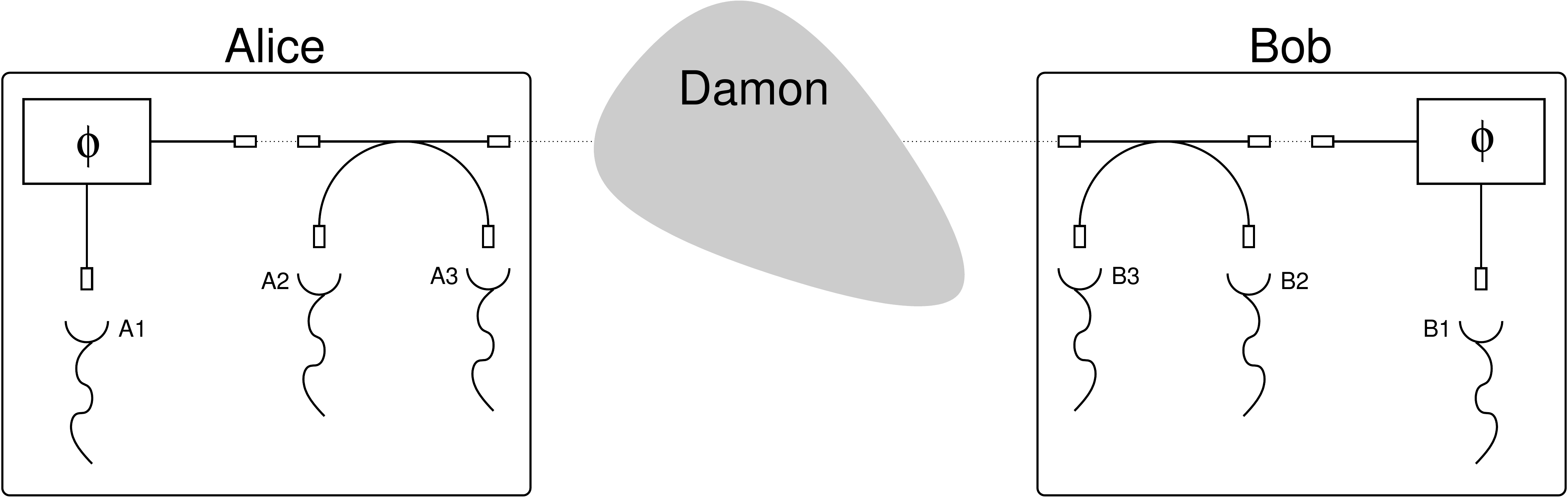}
\caption{A clock synchronization scenario using entangled photons. Alice and Bob each have a source of polarization entangled photon pairs $\Phi$. Alice detects and time tags one member of the pair locally in a particular polarization basis and sends the other member of the pair to Bob, who also time tags and measures his choice of polarization. By measuring joint probabilities of the different polarization combinations, Alice and Bob can determine the value of a Bell parameter and estimate if any geometric phase has been added to the propagating photon. The same procedure is also performed for photon pairs originating at Bob's side.} \label{fig:clock_attack}
\end{figure}

In this section we will look at how this geometric phase can be used to detect a channels non-reciprocity in the presence of a malicious adversary. We base the analysis on a hypothetical twin circulator attack as in Fig.~(\ref{fig:clock_attack}) on the clock synchronization protocol from \cite{lamas-linares:18}, and in particular consider the coincidence probability for Alice's and Bob's polarization measurements on Bell pairs originating at Alice's station.  Thus, the photons measured by Bob have traveled through the channel containing Damon's circulator attack, while the photons that Alice detects undergo no evolution of the polarization state. 

For concreteness, we will assume that the FRs in Damon's attack device are aligned so that they rotate the polarization on the linear polarization plane, i.e. a full rotation around the axis defined by the circular polarization basis.  This assumption is not required, but it will make the analysis below a bit less tedious.  From the discussion earlier, we know that a photon in the state
\begin{equation}\label{eq:psi}
\ket{\psi} = \cos(\theta/2)\ket{R} + \sin(\theta/2)\ket{L}, 
\end{equation}
will acquire an overall phase of $\beta = -\frac{1}{2}\Omega = -\pi(1-\cos\theta)$ due to this full rotation.    
Also, if the photon is in the orthogonal state
\begin{equation}\label{eq:psi_perp}
\ket{\psi_\perp} = -\sin(\theta/2)\ket{R} + \cos(\theta/2)\ket{L}, 
\end{equation}
then the induced geometric phase is $-\beta = +\pi(1-\cos\theta)$.  

Let the entangled pair initially be in the Bell state $\ket{\Phi_-} = \frac{1}{\sqrt{2}}\left( \ket{HV} - \ket{VH}\right)$.  With the first qubit Alice's photon and the second one Bob's photon.  We can re-write the Bell state in the basis defined by  Equations \ref{eq:psi} and \ref{eq:psi_perp},
\begin{align}
\ket{\Phi_-} & =\frac{1}{\sqrt{2}} \Big( \ket{HV} - \ket{VH} \Big) \nonumber \\
 &= \frac{i}{\sqrt{2}} \Big( \ket{\psi_\perp}_A\ket{\psi}_B - \ket{\psi}_A\ket{\psi_\perp}_B \Big). 
\end{align}
The state of the Bell pair after Bob's photon goes through Damon's circulator based attack, $\hat{U}_{Attack}$, is given by 
\begin{align}
\ket{\tilde{\Phi}_-} & = \hat{U}_{Attack}\frac{i}{\sqrt{2}}\Big( \ket{\psi_\perp}_A\ket{\psi}_B - \ket{\psi}_A\ket{\psi_\perp}_B \Big) \nonumber\\
&= \frac{i}{\sqrt{2}} \Big( e^{-i\beta}\ket{\psi_\perp}_A\ket{\psi}_B - e^{+i\beta}\ket{\psi}_A\ket{\psi_\perp}_B \Big). \nonumber
\end{align}
We can see from this expression that the geometric phase induced by Damon's attack has rotated the full Bell state, and this transformation is \emph{not} local to either of the photons.  This rotation will have no effect on the local outcome probabilities for Alice and Bob.  However, as we will show below, this rotation will have a measurable effect on the coincidence probabilities for Alice's and Bob's polarization measurements.

Alice will measure her photon in the linear polarization basis defined by 
\begin{align}
\ket{\alpha}_A &= \cos(\alpha/2)\ket{H}_A + \sin(\alpha/2)\ket{V}_A \nonumber\\
\ket{\alpha_\perp}_A &= -\sin(\alpha/2)\ket{H}_A + \cos(\alpha/2)\ket{V}_A. \nonumber
\end{align}
And Bob will measure his photon in the basis defined by
\begin{align}
\ket{\theta}_B &= \frac{1}{\sqrt{2}} \big( \ket{\psi}_B + \ket{\psi_\perp}_B \big) \nonumber\\
&= \cos(\theta/2)\ket{V}_B - i\sin(\theta/2)\ket{H}_B \nonumber\\
\ket{\theta_\perp}_B &= \frac{1}{\sqrt{2}} \big( \ket{\psi}_B - \ket{\psi_\perp}_B \big). \nonumber
\end{align}
We now ask: What is the probability of the outcome $\ket{\alpha}_A \ket{\theta}_B$?  
\begin{align}
\textrm{Prob}\big( \ket{\alpha}_A \ket{\theta}_B \big) = \left|\bra{\alpha}\braket{\theta}{\tilde{\Phi}} \right|^2. \nonumber
\end{align}
The overlap between the post-selected state and the rotated Bell state is 
\begin{equation}
\begin{aligned}
\bra{\alpha}\braket{\theta}{\tilde{\Phi}}  = {} & \frac{i}{2} \Big \{ {}_A\bra{\alpha} \big( {}_B\bra{\psi} + {}_B\bra{\psi_\perp}\big) \nonumber\\
& \big( e^{-i\beta}\ket{\psi_\perp}_A \ket{\psi}_B - e^{+i\beta} \ket{\psi}_A \ket{\psi_\perp}_B\big) \Big \} \nonumber \\
\end{aligned}
\end{equation}
\begin{align}
\bra{\alpha}\braket{\theta}{\tilde{\Phi}} &= \frac{i}{2} \left[ \bra{\alpha}\psi_\perp\rangle e^{-i\beta} - \braket{\alpha}{\psi} e^{+i\beta}\right] \nonumber\\
&= \frac{i}{2\sqrt{2}} \Big \{ e^{-i\beta} \bra{\alpha}\big( -\sin(\theta/2)\ket{R}+ \cos(\theta/2) \ket{L} \big) \nonumber \\ 
&\quad - e^{+i\beta} \bra{\alpha}\left( \cos(\theta/2)\ket{R} + \sin(\theta/2)|L\rangle \right) \Big \} \nonumber \\
&= \frac{i}{\sqrt{2}} \Big \{ \sin\beta \cos(\theta/2) \sin(\alpha/2) \nonumber \\
& \quad  -i \cos\beta \sin(\theta/2) \sin(\alpha/2) \nonumber \\
& \quad + \cos\beta \cos(\theta/2) \cos(\alpha/2) \nonumber \\
& \quad -i  \sin\beta \sin(\theta/2) \cos(\alpha/2) \Big \} \nonumber
\end{align}
Taking the square modulus of this, we obtain
\begin{align}
\left|\bra{\alpha}\braket{\theta}{\tilde{\Phi}} \right|^2 &=\frac{1}{4} \Big \{ \big( \sin^2(\theta/2 + \beta) + \sin^2(\theta/2 - \beta) \big) \sin^2(\alpha/2)  \nonumber \\
& \quad + \big( \cos^2(\theta/2 + \beta) + \cos^2(\theta/2 - \beta) \big) \cos^2(\alpha/2) \nonumber \\ 
& \quad + \sin(2\beta) \sin(\alpha) \Big \} \nonumber
\end{align}

If we set $\theta=0$, then Bob's measurement is in the $\{V,H\}$ basis since $|\theta\rangle = |V\rangle$, and so both Alice and Bob are measuring in the linear polarization plane.  Note also that $\beta = -\pi(1-\cos\theta)=0$ for the configuration described in the first section where the circulators rotate the polarization about the $\{R,L\}$.  The joint probability then simplifies to 
\begin{equation}
\left|\bra{\alpha}\braket{\theta=0}{\tilde{\Phi}} \right|^2 = \frac{1}{2} \cos^2(\alpha/2) = \frac{1}{4}(1+\cos\alpha).
\end{equation}
From this result, we can see that when the plane defined by Alice's and Bob's measurements is the same plane containing the path induced by the circulators (linear polarization in this case), then the outcome probabilities are independent of the geometric phase $\beta$. In this orientation of the measurement and circulator path planes, the geometric phase becomes an \emph{overall} phase for both Alice's and Bob's qubits and is therefore unobservable.

\begin{figure} 
\includegraphics[width=1.1\linewidth]{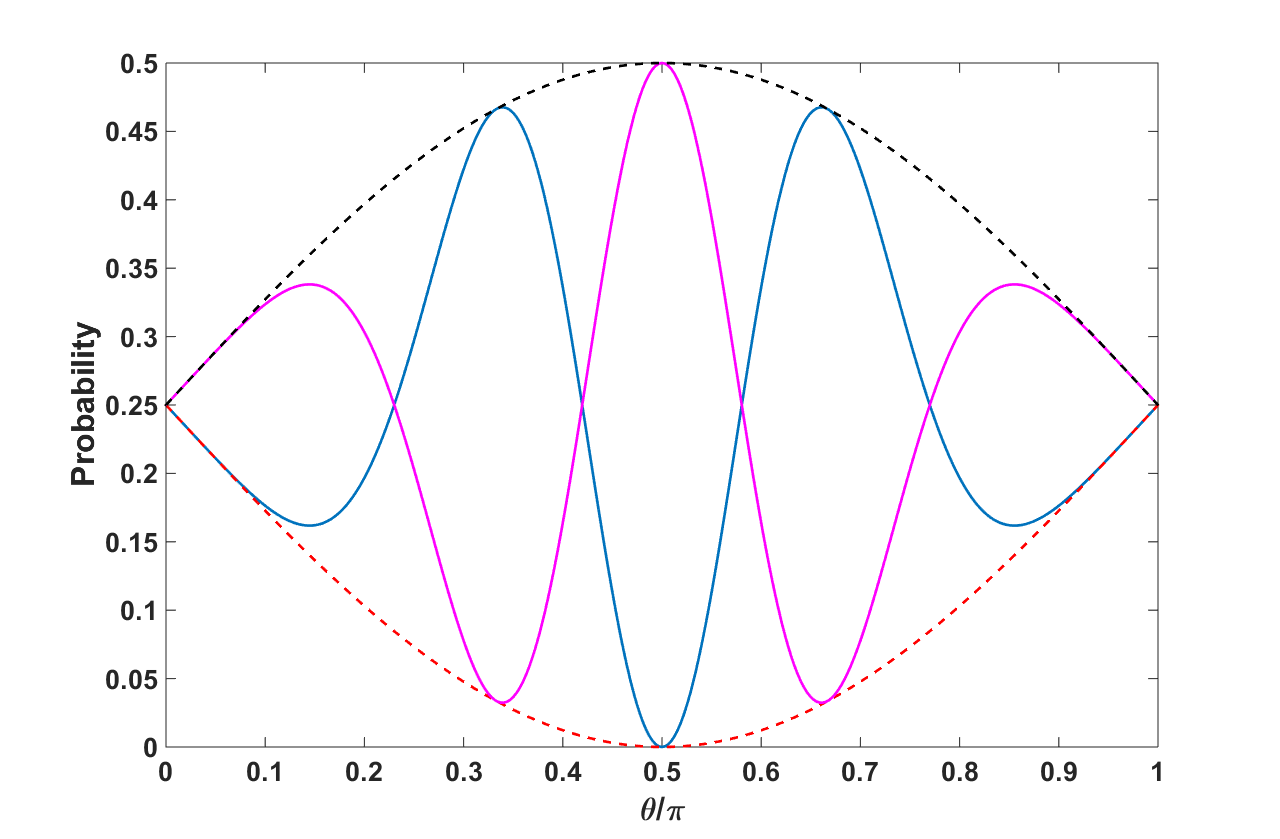}
\caption{Solid curves: The outcome probabilities for the post-selected states $|R\rangle_{Alice}|\theta\rangle_{Bob}$ (Blue) and $|R\rangle_{Alice}|\theta_\perp\rangle_{Bob}$ (Magenta) as a function of $\theta$ when the induced geometric phase is $\beta = \pm\pi(1-\cos\theta)$. Dashed curves: The predicted outcome probabilities for each post-selected state in the absence of a geometric phase.}\label{fig:Prob}
\end{figure}

However, if Alice and Bob measure in a plane perpendicular to the plane defined by the circulators' path, we see something very different. Fixing Alice's measurement basis to be the circular basis $\{R,L\}$, the outcome probabilities as a function of Bob's basis defined by $\theta$ are 
\begin{align}
\left|\bra{R}\braket{\theta}{\tilde{\Phi}} \right|^2 &= \frac{1}{4} \big( 1 + \sin\theta \cos(2\beta) \big), \label{eq:theta} \\
\left|\bra{R}\braket{\theta_\perp}{\tilde{\Phi}} \right|^2 &= \frac{1}{4} \big( 1 - \sin\theta \cos(2\beta) \big) \label{eq:thetaPerp} .
\end{align}
In the case of no geometric phase, $\beta=0$, then $|\tilde{\Phi}\rangle = |\Phi\rangle$.  We can see from Equations \ref{eq:theta} \& \ref{eq:thetaPerp} that the presence of a non-reciprocal device in the channel from Alice to Bob that utilizes a Faraday rotator (a therefore necessarily produces a non-zero $\beta$) will have a measurable effect on the joint outcome probabilities.  In Figure~\ref{fig:Prob} we compare the outcome probabilities with and without the nonreciprocal device in the channel. 

Importantly, the effect of the geometric phase on the outcome probabilities depends on the orientation of the plane defined by Alice's and Bob's measurement bases (assuming they are distinct) and its relation to the rotation plane defined by the non-reciprocal device.  This means that Damon cannot compensate for the induced geometric phase without knowing in advance the bases in which both Alice and Bob will measure for each entangled photon pair.  This is a consequence of the fact that the relevant phase is a nonlocal one defined between Alice's and Bob's qubits. The geometric phase in the quantum case is not gradually built up locally by Bob's photon as it passes through the circulators. This is a fundamental distinction between a truly quantum geometric phase, e.g. the Aharonov-Bohm effect \citep{ab:59,aharonov:16}, and any analogous classical geometric phase, for example the rotation of the plane of oscillation of a Foucault pendulum over the course of one day \cite{bergmann:07}. An interesting implication of the distinction between quantum and classical geometric phase is that reliable detection of non-reciprocity in an adversarial context, such as the secure clock synchronization protocol in \cite{lamas-linares:18}, can only be accomplished by using entangled photon pairs, since while classical light will experience an induced geometric phase, this phase is locally measurable by an adversary and can therefore be compensated.

\section{Conclusions}
The presence of a non-reciprocal device in an optical channel introduces a non-trivial geometric phase in the polarization of the propagating light. When the light undergoing this phase change is a photon that is part of an entangled photon pair, we show that this results in a non-local phase that can be detected by joint probability measurements. Critically, this phase change cannot be ``compensated away'' by a malicious party with the objective of hiding the non-reciprocal nature of the channel. This result has applications in secure metrology and in the understanding of non-reciprocal tools in quantum information technologies.

\bibliography{clock}

\end{document}